\documentclass[letterpaper,twocolumn,10pt]{article}
\usepackage{usenix,epsfig,endnotes}

\usepackage{xcolor}
\usepackage{xspace}
\usepackage{array}
\usepackage{amsmath}
\usepackage{tabularx}
\microtypecontext{spacing=nonfrench}
\newcolumntype{L}[1]{>{\raggedright\arraybackslash}m{#1}} 
\usepackage{hyperref}
\usepackage[usestackEOL]{stackengine}
\usepackage{multirow}
\usepackage{tablefootnote}
\begin{document}
\newcommand{\name}{Clid\xspace}
\newcommand{\todo}[1]{\textcolor{blue}{#1}}

\date{}

\title{\Large \bf Clid: Identifying TLS Clients With Unsupervised Learning on Domain Names}

\author{
{\rm Ihyun Nam}\\
Stanford University
\and
{\rm Gerry Wan}\\
Stanford University
}

\maketitle

\thispagestyle{empty}


\section*{Abstract}
In this paper, we introduce \name, a Transport Layer Security (TLS) client identification tool based on unsupervised learning on domain names from the server name indication (SNI) field. \name aims to provide \textit{some} information on a wide range of clients, even though it may not be able to identify a definitive characteristic about each one of the clients. This is a different approach from that of many existing rule-based client identification tools that rely on hardcoded databases to identify granular characteristics of a few clients. Often times, these tools can identify only a small number of clients in a real-world network as their databases grow outdated, which motivates an alternative approach like \name.

For this research, we utilize some 345 million anonymized TLS handshakes collected from a large university campus network. From each handshake, we create a TCP fingerprint – comprising IP flags, time-to-live (TTL), TCP window size, initial sequence number, window size, flags, header length, options, max segment size, and window scaling – that identifies each unique client that corresponds to a physical device on the network. \name uses Bayesian optimization to find the optimal (in a precise sense that we define later) Density-Based Spatial Clustering of Applications with Noise (DBSCAN) clustering of clients and domain names for a set of TLS connections. \name maps each client cluster to one or more domain clusters that are most strongly associated with it based on the frequency and exclusivity of their TLS connections. While learning highly associated domain names of a client may not immediately tell us specific characteristics of the client like its the operating system, manufacturer, or TLS configuration, it may serve as a strong first step to doing so. There exists prior work \cite{sni, sni1} that uses the SNI field for client identification.

We evaluate \name's performance on various subsets of our captured TLS handshakes and on different parameter settings that affect the granularity of identification results. Our experiments show that \name is able to identify the single most associated domain cluster (a group of \textit{similar} domain names in a precise sense that we define in \S\ref{sec:distance}) for at most 90\% of clients in 10,000 TLS connections for a real-world traffic. When one or more domain clusters were allowed to be mapped to a single client cluster, \name identified such domain names for at least 60\% of all clients in all our experiments.

\section{Introduction}
Identifying clients in network communications is widely used for various research purposes, such as studying anomalous behaviors on the network. However, many current client identification are instead of identifying a wide range of clients accurately.

We introduce \name, a passive TLS client identification tool. There is often a trade-off between identifying something definitive about a client and getting some information about many clients, even though it is not very granular. Many state-of-the-art client identification tools look for a definitive identification of clients by establishing temporary ground truth in the form of databases that act as guidelines or simple dictionaries to identify clients \cite{Lastovicka, zardaxt, p0f}. However, solutions that rely on databases can become unreliable as its database grows outdated. \name takes a different approach and instead tries to identify as many clients as possible, even though it may not be able to say something definitive about each one of them. \name does so by finding a mapping between a client and its associated domain names without relying on a database. Based on the selected domain names, a user may learn some information about the client's operating system, manufacturer, IoT device type, and more.

In our experiments using 10,000 anonymized TLS connections from real-life traffic, Clid was able to identify the single most strongly associated domain name for at most 90\% of clients, based on our criteria for client-domain association. For at least 60\% of clients in all our experiments on varying numbers of TLS connections, ranging from 2,000 to 10,000, Clid identified the most associated domain names with varying degrees of association with the clients.
\section{Background} \label{sec:background}
Client identification is the practice of identifying traits of a client involved in an online communication, such as its operating system (OS), TLS configurations, or hardware specifications. Client identification is a widely used methodology for researchers and operators because it enables them to learn about a particular group of clients and specify their behaviors. For example, client identification can help researchers identify devices that may be causing abnormal behaviors. In this work, we focus on passive client identification, which aims to identify clients through passive observation of network traffic, as opposed to directly engaging in communications. A client identification tool typically receives recorded or live traffic as input and extracts some target properties that are observable in the traffic. Three of the most common properties used in client identification are: the client's Transmission Control Protocol and Internet Protocol (TCP/IP) parameters, user-agent, and the server name indication (SNI) field \cite{survey}. A typical client identification tool then processes these information to return a classification result, which can be the OS name and version, manufacturer, device type, or something else about the client.

\section{Motivation} \label{sec:motivation}
Many previous client identification methods rely on creating databases that act as a simple dictionary of network parameters and client types \cite{Lastovicka, zardaxt, p0f}. However, databases risk growing outdated. In fact, the inefficiency of such rule-based fingerprinting tools in classifying clients has long been known to the academic community~\cite{Lastovicka}. 
We see through a simple experiment on some of the most well-known OS fingerprinting databases, namely Joy~\cite{joy}, p0f\cite{p0f}, and Zardaxt\cite{zardaxt}, that databases have many missing values. We used Retina to collect anonymized copies of all TLS handshake transcripts sent over a large university campus network over a three week period in 2022. We observed about 550M handshakes per day and collected, among other things, the TCP fingerprint of SYN packet of each handshake, which comprises IP flags, time-to-live (TTL), TCP window size, initial sequence number, window size, flags, header length, options, max segment size, and window scaling. Among 2000 of the anonymized TCP fingerprints, a maximum of 50.1\% of them appeared in any one of the databases and an average of 13.6\% of them appear in any two of the three databases, while only 5.85\% of them appear in all three databases. We see that the three databases together can (1) identify very few clients and (2) even when they all identify a client, they rarely agree. As we see in table \hyperref[tab:dbpercent]{1} only 3.05\% of our TCP fingerprints are identified as the same OS by all three databases, and at most 10.95\% by any two databases. In other words, around 97\% of the time, at least one of Joy, p0f, and Zardaxt mis-identifies a client.

\begin{table*}[ht!]
\label{tab:dbpercent}
    \centering
    \begin{tabular}{|c|c|c|} 
    \hline
    Database names & \% of clients that appear in all databases & \% of clients identified as the same OS \\
    \hline
    Joy, Zardaxt, p0f &  5.85 & 3.05 \\
    Joy, Zardaxt  & 16.55 & 0\\
    Joy, p0f & 5.85 & 2.50\\
    Zardaxt, p0f & 18.50 & 10.95 \\
    \hline
    \end{tabular}
    \caption{Identification results from different databases}
\end{table*}

It also takes considerable administrative effort to maintain the created databases up-to-date and Joy, p0f, and Zardaxt are seemingly not updated automatically. This is why many existing OS fingerprint databases are abandoned; for example, p0f was last updated in 2016 \cite{p0fgithub}. The landscape of the modern network changes rapidly today as the market shares of relatively new OSes like Mac OS and iOS grow \cite{os-market} and new IoTs emerge rapidly \cite{iot-market}. This makes databases grow outdated quickly and under-representative of many clients in real networks, especially those that carry traffic from relatively new operating systems like iOS and Mac OS. For example, Joy's database, which was made public in 2019, only contains fingerprints for Windows, Mac OS, and Linux. Similarly, p0f's database contains 1 iOS, 4 Mac OS, 6 Windows, and 17 Linux fingerprints. It may also be the case that these databases focused on collecting certain types of devices and thus not a wide range of devices was included. We see that a novel tool that does not rely on databases is needed to provide more accurate and reliable information about clients' OS and further, more characteristics of TLS clients.

More recent client fingerprinting tools use supervised learning like machine learning to train tools that can identify certain clients. However, a major drawback of these tools is that they are trained on precisely the set of clients they aim to identify, and therefore their capability often ends at discerning the target clients from all others \cite{survey}. With \name, we get a step closer to learning meaningful information about any unspecified client, which is useful when analyzing an unknown network.


What makes it challenging to evaluate the accuracy of passive client identification tools is the absence of ground truth on the identities of the client. Given that attempting to establish ground truth with databases is highly inefficient, an alternative solution is to find associations between available parameters in TLS connections and their clients intelligently. SNI is an unencrypted field in the client hello of a TLS handshake that provides a good window into the client's identity. SNI has been used as part of several previous client identification tools \cite{Lastovicka, sni1}. Furthermore, previous research efforts have shown that SNI offers good insight into the type of service offered by the website \cite{sni}, which enables us to envision what kind of clients would access such sites. Nonetheless, most previous efforts that use SNI establish a dictionary of a small subset of domain names known to be accessed by certain types of clients and rely on a simple mapping to identify clients \cite{Lastovicka}.

Without relying on a database, a strawman approach way to use SNI for client identification is for a human to manually inspect all SNIs that a client connected to and infer an identity for the client. For example, it may be reasonable to conclude that a device that repeatedly and frequently connects to \texttt{login.apple.com}, \texttt{icloud.com}, and \texttt{apple.update.com} is likely an Apple product. But to do so, a person would have to go through the hundreds of domain names that the device connected to and be able to understand which domain names are informative. This is not only inefficient but also scientifically uninteresting. \name attempts to automate this process of extracting client identity from SNIs and imitate a human's learning process using machine learning.

\section{Design Goals}\label{sec:goal}
We aim to build a tool that can be used by researchers and other real-life users to learn more about their TLS clients. In this section, we introduce a set of design goals that we aim to achieve with \name. We believe these goals will help us overcome the shortcomings of previous client identification tools. \name does not aim to definitively claim a client to be a particular something. Instead, \name identifies a strong association between clients and some domain names they connect to, which serves as a first step in identifying the clients.

\textbf{Breadth over granularity of identification.} Many traditional client identification tools can provide very detailed (or 'granular') identification of clients. For example, there exist tools that can identify the major and minor OS versions of clients \cite{osminor, Lastovicka, osminor1} and tools that can tell the exact model and manufacturer of particular IoTs \cite{device}. Some tools achieve such granularity by training a machine learning model on the test data itself through supervised learning. This might limit the accuracy of these tools to only a small subset of clients that they are trained on. Other tools create a dictionary after establishing ground truth and identify clients through simple mapping. However, such databases risk growing outdated quickly as the types of devices in networks evolve. \name takes an alternative approach; name must prioritize breadth over granularity of client identification so it can provide some information, even though not granular, on more clients than the state-of-the-art solutions. For example, \name need not be able to tell a user that a client is an iPhone 13 with iOS version 17.0.1; it is sufficient to tell that it is likely an iPhone. Despite this trade-off, we believe \name wil make a useful tool for researchers and operators because a rough identification of devices still suffices many research and operational purposes \cite{rough1, rough2}.

\textbf{Give more information about each client.} Most previous client identification techniques are designed to identify one aspect of the clients, such as their OS \cite{joy, p0f, zardaxt, latovicka2}, user-agent \cite{ua}, version and model of IoT devices \cite{iot1, iot2, iot3}, or the manufacturer of wireless devices \cite{manu1, manu2}. Such targeted identification can be helpful when the user has a specific learning objective in mind or knows what types of clients to expect in the network traffic to be analyzed. However, a more general identification tool is needed when the user does not have such information or the traffic to analyze contains various types of clients; for example, some may be running on an OS while some may not be, or some might be an IoT while some are not. To help with these cases, \name must be able to provide a large variety of information about clients, including but not limited to OS, type of device, or IoT type that certain domain names may suggest. A user might not learn all those things about one client.

\textbf{Performance does not degrade.} Most previous approaches at client identification \cite{ua, db1, Lastovicka, db2} establish some ground truth, usually in the form of a database, that maps observable properties to clients. They then search for either exact or partial matches in the database to classify clients. However, in order not to degrade over time, \name must not rely on hardcoded databases or supervised learning.
\section{Design Overview} \label{def:main}
In this section, we provide an overview of \name's design and explain its main modules in detail. Furthermore, we report results for micro-benchmarks to show how the design choices made satisfy the goals identified in \S\ref{sec:goal}. \name first uses Bayesian optimization and DBSCAN clustering to group clients and domain names, separately, based on similarity metrics we define in \S\ref{sec:distance}. \name then uses unsupervised learning and concepts from bipartite graphs to map each client cluster to one or more domain clusters that are deemed most strongly associated (in the precise sense that we define as \textit{weight} in \S\ref{sec:weight}) with the identify of the client. The Python source code for \name and experiment results are available on our Github page.\footnote{https://github.com/ihyunnam/clid}

\name takes as input a set of TLS connections; they must have the SNI field populated to be used by \name. For each TLS connection, \name generates a \textit{TCP fingerprint} by concatenating the values of its TCP header length, IP time-to-live, TCP window size, TCP flags, TCP maximum segment size, TCP options, and TCP window scaling parameters. Any unpopulated field is replaced with an empty string. While these parameters are known to be influenced by the software and hardware of the client device \cite{param1, param2}, none of them on its own can be strongly indicative of the physical device. However, considering all seven parameters at the same time creates a strong association to a physical device on a network\cite{tcp1, tcp2, tcp3}. Therefore, we use TCP fingerprints as unique identifiers for clients. The input set of TLS connections may contain multiple connections requested by the same client, in which case \name generates multiple copies of the same TCP fingerprint and retains all of them in a list. It becomes important that \name does this instead of recording only distinct TCP fingerprints when computing weights between client and domain clusters, as we will see in \S\ref{sec:weight}.

For each TLS connection in the input database, \name also extracts the SNI field value ("domain") and stores all of them in one list where duplicates are allowed. \name performs DBSCAN clustering on these lists of clients and domains, separately, using the optimal epsilon values chosen through Bayesian optimization in \S\ref{sec:bayesopt}.

For each client cluster $c$, \name loops through all domain clusters and computes the weight $W_{c,d}$ between $c$ and each domain cluster $d$ using our weight formula. This construction is analogous to building a connected bipartite graph with weighted edges where the client clusters and domain clusters form two sets of nodes. The higher the weight, the more indicative \name thinks $d$ is of $c$'s identity. After computing all weights, \name returns the highest weighed domain cluster(s) for each client cluster. A client cluster is said to be \textit{mapped} to the domain cluster(s) that was given the highest weight.

A user can then manually inspect the domains included in the chosen domain cluster and infer what they tell about the clients. For example, if a client cluster is mapped to a domain cluster containing various paths to \texttt{icloud.com}, it is reasonable for the user to infer that the client is likely an Apple product.

This way, \name extracts only the meaningful domain names that each client connected to, while discarding irrelevant domain names that are not strongly associated with the client. This partly automates the process of a human going through all domain names a client connected to, which may contain irrelevant or misleading information, and deducing the client's identity from them.

\subsection{Maximizing the Number of Good Client Clusters}
\label{sec:bayesopt}
In this subsection, we introduce the concept of a \textit{good} client cluster and explain how \name uses it as a metric of success for Bayesian optimization for client clustering.

A good client cluster is defined with regards to two parameters: \texttt{Z} and \texttt{H}. In order to be good, a client cluster needs to be mapped to a domain cluster whose weight has a minimum z-score of \texttt{Z}. Z-score is a common measurement of distance between data points in statistics, which indicates how many standard deviations a particular point is away from the mean of all data points. In this work, \texttt{Z} is computed as
\[\texttt{Z} = \frac{W-\mu}{\sigma}\]
for $W$, the highest weight a client cluster has with any one domain cluster; $\mu$, the mean of all weights a client cluster gets; and $\sigma$, the standard deviation of all those weights. For example, if $\texttt{Z}=5$, a client cluster needs to be mapped to a domain cluster with weight at least five standard deviations higher than the mean of all weights in order to be good. On the other hand, if $\texttt{Z}=1$, a good client cluster only needs to have the highest weight domain cluster be weighed at least one standard deviation higher than the mean. In this sense, \textit{good-ness} is a measure of confidence in the result.

\texttt{H} is the maximum number of highest-weight domain clusters a good client cluster may be mapped to. For example, if $\texttt{H}=2$, \name need not pinpoint a single domain cluster to be highly associated with a client cluster, in order to make it good. The client cluster can be mapped to \textit{up to} 2 domain clusters whose weights are both \texttt{Z} standard deviations higher than the mean.

We now explain how Bayesian optimization uses the number of good client clusters as a metric of success to choose the optimal epsilon values for DBSCAN clustering. The epsilon value in DBSCAN clustering is the maximum distance allowed between the core point of a cluster (selected randomly at the beginning) and any other point belonging to the cluster. The other input to DBSCAN clustering, \texttt{min\_samples}, which specifies the minimum number of data points in a cluster, is fixed at 1 to allow for singleton clusters. We choose Bayesian optimization over other optimization algorithms due to its ability to compute expensive evaluation functions and to intelligently pick the inputs for the next iteration based on previous iterations.

\begin{figure}[h!]
    \centering
    \includegraphics[width=0.5\linewidth]{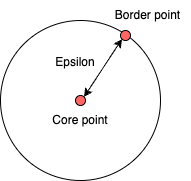}
    \caption{Structure of a DBSCAN cluster}
    \label{fig:enter-label}
\end{figure}

Bayesian optimization receives as inputs an epsilon value for client clusters and an epsilon value for domain clusters, both in range (0,1). Because we normalize the distance between data points (clients and domain names, separately) to the [0,1] (inclusive) range, setting the allowed epsilon values to (0,1) (exclusive) ensures that the two farthest data points, with distance equals to 1, in any database are not clustered together. The evaluation function of Bayesian optimization aims to maximize both the number of good client clusters and the number of domain clusters formed as a result of DBSCAN clustering. While only the number of good clusters is an official metric of success for \name's mapping, the number of domain clusters is added to the return value to discourage DBSCAN clustering from resorting to a trivial solution in which it clusters all domain clusters together and maps every single client cluster to it. In this trivial solution, all client clusters are marked as good, as they are all mapped to a single, non-zero weight domain cluster. However, this behavior must be discouraged, because it fails to identify meaningful domain names for each client.

After 10 iterations of Bayesian optimization, the optimal epsilon values for clients and domains are determined. DBSCAN clustering is done separately on clients and domains with the respective epsilon values to yield the final clusters \name uses for mapping.

\subsection{Calculating Weights}
\label{sec:weight}
Here, we explain how \name computes the weight between a client cluster and a domain cluster. Furthermore, we report results for micro-benchmarks to justify our weight formula. Consider a client cluster $C$ and a domain cluster $D$. We denote the weight between $C$ and $D$ as $W_{C,D}$ and compute it as
 \[W_{C,D} = \frac{f_{C,D}}{e_{C,D}}.\]
Here, $f_{C,D}$ denotes the frequency, and $e_{C,D}$ denotes the non-exclusivity between $C$ and $D$. Frequency and non-exclusivity are defined more precisely below.

\[f_{C,D} = \frac{\sum_{c\in C} \text{ number of } d\in D \text{ that } c \text{ connected to}}{|C|}\]

\[e_{C,D}= \frac{\sum_{d \in D} \text{number of connections $d$ made with $c \notin C$}}{|D|} \]

We see that $W_{C,D}$ is proportional to the average number of times the clients in $C$ (denoted $c\in C)$ connects to the domains in $D$ (denoted $d\in D$), and inversely proportional to the average number of times some $d\in D$ appears in all other client clusters that are not $C$ (denoted $C'$). That is, for $W_{C,D}$ to be high, (1) $d\in D$ has to be frequently accessed by $c\in C$ as opposed to a one-time connection, and (2) $d\in D$ has to be connected exclusively by $c\in C$ and as few times as possible by clients in other client clusters ($c'\in C'$).

The frequency factor $f_{C,D}$ ensures condition (1) by giving lower weights to non-repetitive domain connections. Domain names should appear multiple times a client's TLS connections in order for it to be considered representative of the client's identity. For example, a client connecting to \texttt{weather.apple.com} might appear to be an Apple device, but if the connection happens once among a hundred connections, the client could also be an Android device checking the weather on a browser. When connections are not frequent, we cannot reasonably infer the client's identity. Condition (2) is represented by the non-exclusivity factor $e_{C,D}$ in the weight formula. It is needed because there are common domains like those of email providers and streaming services that can be accessed by all devices regardless of their specifications. The non-exclusivity factor works against the frequency factor to discourage domains from being weighed high simply due to the sheer number of connections made, when in reality, for example, all other client clusters connected to the domain equally many times. Because \name creates a list of \textit{all} occurrences of domains in the input TLS database, as opposed to keeping a single copy of each distinct domain, it is able to check exactly how many times a domain was accessed by clients in each cluster.

\subsubsection{The Importance of Frequency} \label{sec:freq}
This micro-benchmark serves as a justification for considering frequency in the weight formula. We remove frequency from the weight formula so that the new formula becomes 
\[W_{C,D} = \frac{1}{e_{C,D}}.\]
We tested \name to cluster 2,000 TLS connections using the new weight formula. Both \texttt{Z} and \texttt{H} values were set to 1. As a result, \name managed to make 100\% of client clusters good, mapping every client cluster to one domain cluster. However, this result is misleading, because the chosen domain clusters were often made of generic domains like \texttt{linkedin.com} and \texttt{cloudfront.com} that tell us little about the client. This could happen because non-exclusivity is computed with respect to connections that other client clusters made in the database.

As a demonstrating example, Table \hyperref[tab:freq]{2} shows the mapping outcome from this experiment for a particular client cluster (call this \texttt{C}) that connected to seven different domains. The seven domains were clustered into five distinct domain clusters as shown in the table. A human may reasonable tell from inspecting the domains of domain cluster 5 that the clients in \texttt{C} are most likely Apple devices, while discarding other domain clusters as they are not indicative of a particular OS, device, or IoT.

\begin{table*}
\label{tab:freq}
    \centering
    \small
    \begin{tabular}{|c|c|c|c|c|c|c|}
    \hline
     \Centerstack{Domain\\ cluster}& Domain name & \Centerstack{Weight with\\no frequency} & \Centerstack{Weight with\\correct formula}& \Centerstack{Number of connections \\to other client clusters} & \Centerstack{Number of connections \\to client C} \\
    \hline
    \hline
    1& linkedin.com & 1.33& 0.19 & 1 & 1\\
    \hline
    2&trustarc.com\tablefootnote{Website to a company that provides data privacy management solutions.} & 2.00 & 0.29& 0 & 1\\
    \hline 
    3&gstatic.com\tablefootnote{A website owned by Google that helps the contents for Google services load faster from Google servers.} & 0.14 & 0.15& 7 &1\\
    \hline 
    4&cloudfront.net \tablefootnote{A website to Amazon's Cloudfront, a content delivery network for Amazon.} & 0.14& 0.16&3&1\\
    \hline
    5&{\Centerstack{(3 different paths to) apple.com}} & 1.01& 0.43& 315&3\\
    \hline
    \end{tabular}
    \caption{Mapping results for the same client cluster using a weight formula with and without the frequency factor. We are only displaying the second- and top-level domains.}
    \label{tab:my_label}
\end{table*}

when using the weight formula \textit{without} the frequency factor, \texttt{C} was mapped to domain cluster 2 that contains \texttt{choices-or.trustarc.com}, because it has zero connections to other devices. \name is unable to take into account the fact that this domain was accessed by \texttt{C} only once. On the other hand, using the correct weight formula \textit{with} the frequency factor, \texttt{C} was mapped to domain cluster 5 containing various paths to \texttt{apple.com}. Although these domains have 315 connections to clients that are not in \texttt{C}, clients in \texttt{C} connected 3 times more to these domains than they did to other domains. Therefore, domain cluster 5 is given a high frequency score to complement for its low non-exclusive score and is mapped to \texttt{C}, which aligns with what a human may conclude.

\subsubsection{The Importance of Non-Exclusivity}
This micro-benchmark serves as a justification for considering non-exclusivity in the weight formula. We removed non-exclusivity from the weight formula and tested \name on 2,000 TLS connections with both the \texttt{Z} \texttt{H} values set to 1. The new weight formula is 

\[W_{C,D} = f_{C,D}.\]

Using the new weight formula, 79.57\% of all client clusters were good, which is 10.75\% lower than the percentage of good clusters formed with the original weight formula. This result combined with that from \S \ref{sec:freq} justifies our design of the weight formula.

\subsection{Distance Function for Domain Clustering} \label{sec:distance}
In this subsection, we explain our distance function for DBSCAN-clustering domain names, which we use instead of the default Euclidean distance. \name uses Euclidean distance for client clustering, because clients are represented as concatenated strings, and it suffices to detect unweighted difference between them. However, for domain clustering, we design a more intelligent distance function that takes into account the relative significance of different components of a domain name.

\begin{figure}[ht]
\begin{center}
\includegraphics[width=0.15\paperwidth]{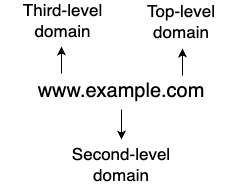}
\caption{Domain name components}
\end{center}
\end{figure}

Our distance function partitions a domain name into components separated by a dot (.) and compares the corresponding strings at the same indices. Any subdomain prefixes such as \texttt{www} are disregarded if present, and the second-level domain is considered to be at index 0. Every time strings mismatch, the distance increases; and the larger the distance, the less likely it becomes that the two domains will be clustered together. We present the formula below.

\vspace{2ex}

\begin{tabularx}{0.9\linewidth}{l@{}c@{}X}
    \text{distance} & = & $\frac{1}{3}$ (if the top-level domains mismatch) \\
                     & + & $\frac{1}{2}$ (if the second-level domains mismatch) \\
                     & + & $(1-0.01s) \frac{1}{i}$ (if ith-level domains have similarity score $s$) \\
\end{tabularx}
\vspace{2ex}

Same domain names will have distance equal to zero. Except for the top-level domain that is given an arbitrarily low weight, more weight is assigned towards the end of the domain. This is because lower-level domains are larger classifications of domain names and therefore carry more information about the website. For example, in \texttt{update.icloud.com}, the distance function perceives \texttt{icloud} more importantly than \texttt{update}, while \texttt{.com} is less important than \texttt{icloud}. Furthermore, components lower than the third-level domain are compared using the fuzzy string matching computed using the Levenshtein distance \cite{Seatgeek} instead of exact matching. This allows the distance function to accommodate for paths that carry similar values but have slightly different names, such as \texttt{profiles-01.example.com} and \texttt{profiles-02.example.com}.

\subsection{Choice of the Clustering Algorithm}
In this subsection, we report results for micro-benchmarks on using \name with different clustering algorithms, and justify our choice of using DBSCAN. We test the following three clustering algorithms.
\begin{itemize}
    \item \textbf{KMeans}: KMeans clustering with default (Euclidean) distance metric for both clients and domains
    \item \textbf{Default DBSCAN}: DBSCAN clustering with default (Euclidean) distance metric for both clients and domains
    \item \textbf{Custom DBSCAN}: DBSCAN clustering with a custom distance metric for domains and default (Euclidean) distance metric for clients
\end{itemize}
KMeans and DBSCAN were considered among many available clustering algorithms because they can perform unsupervised learning on unlabeled data and are compatible with Bayesian optimization. KMeans is used with Bayesian optimization to find the \textit{number} of client clusters and \textit{number} of domain clusters that result in the maximum number of good client clusters. DBSCAN is used with Bayesian optimization to find the ideal \textit{epsilon values} to maximize the number of good client clusters.

We used \name with each clustering algorithm and ran it on 2,000 TLS connections sampled randomly from our database, with \texttt{Z} values 1, 1.5, and 2. We observed that custom DBSCAN that uses our distance function for domain names from \S\ref{sec:distance} yielded the highest number of good client clusters. For all \texttt{Z} values, KMeans resorted to a trivial solution and grouped all clients into two gigantic clusters. Both clusters, however, were not \textit{good}, by our definition of good. Default DBSCAN and Custom DBSCAN were able to make more than 90\% of all clients clusters good for $\texttt{Z}=1$. However, as shown in \hyperref[tab:kmeans]{figure 3}, as \texttt{Z} increased, Custom DBSCAN performed consistently better than Default DBSCAN. Therefore, we chose to use Custom DBSCAN with our domain distance function in \name.

\label{tab:kmeans}
\begin{figure}[ht]
\begin{center}
\fbox{\includegraphics[width=0.4\paperwidth]{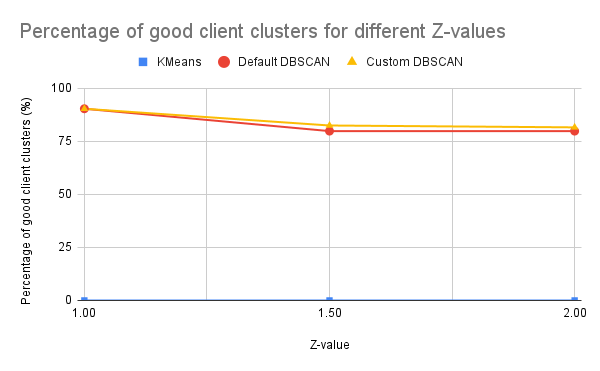}}
\caption{Percentage of good client clusters formed by different clustering algorithms}
\end{center}
\end{figure}
\section{Performance Evaluation} \label{sec:6}
In this section, we report the results of testing \name on various test sets of TLS connections. The metric of success in all tests is the percentage of good client clusters among all client clusters formed. To mimic real-world traffic, all TLS connections were sampled from a database of some 345 million anonymized TLS connections that we observed on a large university campus network in a 24-hour period. All evaluations were done on a c3-highmem-8 machine with 8 vCPUs and 64GB of memory, running on Intel Sapphire Rapids, unless otherwise specified. Most notably, our experiments show that \name can map at least 25\% of all client clusters formed to \texttt{H} number of domain clusters with weight z-score higher than \texttt{Z}, across for \texttt{H} and \texttt{Z} values we tested. The percentage was 50\% when $\texttt{H}=1$.

\subsection{Performance on Different \texttt{H} and \texttt{Z} Values} \label{sec:HZ}
We tested \name on different combinations of \texttt{H} and \texttt{Z} values and measured what percentage of client clusters were \textit{good}. The tested \texttt{H} values are 1, 2, 3, and 4, and the \texttt{Z} values range from 0 to 5, incremented by 0.5 each time. For these experiments, we randomly sampled 1,723 TLS connections that had the SNI field populated from our dataset.

Evaluation results show that for all \texttt{H} values, as \texttt{Z} increases, the percentage of good clusters decreases. This is because with \texttt{Z} value increasing, higher weights are required of domain clusters to make the associated client cluster good. From a usability point of view, a user can have more confidence in the meaningfulness of domain clusters that were mapped with a higher \texttt{Z} value, because that suggests more distinguishability of the chosen domain names from others.

In the trivial case when $\texttt{Z}=0$, all client clusters are marked good as expected. Even for \texttt{Z} as high as 5, 64.52\% of all client clusters were good. Note that, however, this does not mean that seeing the client-domain mapping, a human user would be able to tell definitively the OS, device, or manufacturer of all these clients. Instead, the user is given a set of domain names that have strong enough associations with these clients, which they can then use to infer information about the clients.

Among the good client clusters for when $\texttt{Z}=5$, the majority of clients (86.66\%) only have one connection included in the test set of 2,000 connections to begin with. Therefore, they are each mapped to only one domain cluster with a non-zero weight, which automatically makes them good. This calls for an additional evaluation of \name in \S\ref{sec:diffnum} on clients with different numbers of connections included in the input data set. Furthermore, across all \texttt{Z} values, the higher the \texttt{H} value, the higher the percentage of good clusters.

\begin{figure}
\begin{center}
\fbox{\includegraphics[width=0.4\paperwidth]{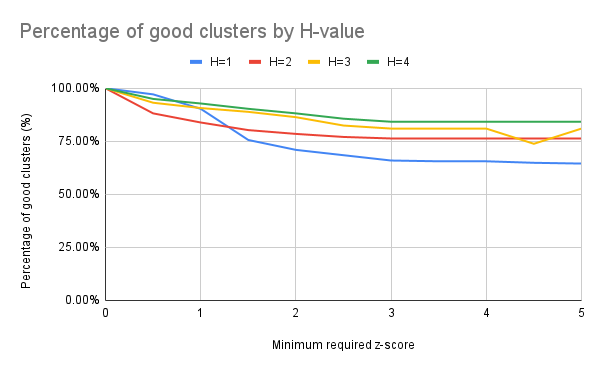}}
\caption{Percentage of good client clusters for different \texttt{H} and \texttt{Z} values}
\end{center}
\end{figure}

\subsection{Identifying Clients With Varying Numbers of Connections}
\label{sec:diffnum}
In this subsection, we aim to test how many connections from a single client \name's unsupervised learning algorithm needs to see in order to map it to a correct domain cluster. From our dataset, we randomly chose a client IP and collected (all) 854 TLS connections issued from it. All 854 connections have the same TCP fingerprint (concatenation of seven TCP/IP parameters as described in \S\ref{def:main}), which convinced us that the client IP represents a single physical device as intended. Call this client \texttt{C}. Among the 854 TLS connections, 214 (or 25.06\%) connections are made to domain names that are identified as belonging to iOS/Mac OS devices in a public database \cite{Lastovicka} that maps a chosen few domain names to the most likely OS of the client based on empirical result. There are also a few domain names, most notably \texttt{updates.apple-cdn.com}, that are not identified by \cite{Lastovicka}'s database but still strongly indicative of an Apple product.
Through the frequent appearance of these domains in connections by \texttt{C}, we establish ground truth that \texttt{Client C} is an Apple device and its OS is either iOS or Mac OS.

We created seven arbitrary data sets containing 2,000 TLS connections each. Each data set included exactly 1, 50, 100, 150, 200, 250, and 300 randomly selected connections made by \texttt{C}, respectively. Both \texttt{Z} and \texttt{H} values were set to 1. 

\begin{table}[ht!]
    \centering
    \begin{tabular}{|c|c|}
    \hline
    Total number of connections by \texttt{Client C} & Good \\
    \hline
    1 &  Yes \\
    50  & Yes  \\
    100 &Yes\\
    150 &Yes \\
    200  & Yes \\
    250& Yes \\
    300 & Yes \\
    \hline
    \end{tabular}
    \caption{Clustering result for \texttt{Client C} with varying numbers of connections}
    \label{tab:numconn}
\end{table}

We see that \name can always identify domain names with the strongest association with \texttt{C}, regardless of how many domain names are available for learning. However, the mapped domain names do not always tell us correctly that \texttt{Client C} is an Apple device. This calls for the notion of meaningfulness of domain names introduced in \S\ref{sec:meaningful}.


\section{Meaningful Domain Names}\label{sec:meaningful}

Ideally, a successful client identification tool should map each client cluster to domain names that are actually informative about the client's identity, in a way that a human manually inspecting all domain names would draw out domain names they think are meaningful. A good client as we define in \S\ref{sec:6} does not necessarily mean that a human user is guaranteed to be able to identify it using the mapped domain names. For example, simply based on our weight formula, a client could be mapped with high confidence to domain names like \texttt{netflix.com}, \texttt{facebook.com} and \texttt{spotify.com}, but these are not informative or meaningful. On the other hand, domain names like \texttt{login-apple.com}, \texttt{update.apple.com}, and \texttt{icloud.com} are \textit{meaningful} because a human user can reasonably infer from these that the client is likely an Apple device.

However, to determine whether a domain name is meaningful, a human user would have to inspect it and decide whether it is informative. This is hard to test. Nonetheless, in \S\ref{sec:meaningful}, we attempted to test how effective \name is in mapping clients to meaningful domain names, among a test set of mixed domain names.

We tested what percentage of TLS connections used in \name's unsupervised learning needs to be \textit{meaningful}, as decided by the authors of this paper, in order for \name to map the client to one of such domain names and not some other domain in the training set. We used the same client \texttt{C} as in \S\ref{sec:6} that we know is an Apple device. We made seven test sets with 2,000 total TLS connections, 300 of which come from \texttt{Client C}. The 300 connections included varying numbers of meaningful domain names: 0\%, 10\%, 20\%, 25\%, 30\%, 40\%, and 50\% \footnote{The SNI-OS database of \cite{Lastovicka} we use to identify meaningful domains is by no means comprehensive, and therefore the actual percentage of meaningful connections for each data set is likely higher than what is intended in the test.}. Table \hyperref[tab:meaningful]{4} shows that \name needs 20\% to 25\% of all connections from a client to be meaningful, in order for it to map the client to those domain names. A desirable next step is to identify how many TLS clients in real life actually meet this criterion and therefore are identifiable with \name.

\begin{table*}[ht!] \label{tab:meaningful}
    \centering
    \begin{tabular}{|c|c|c|c|}
    \hline
    Percentage of meaningful connections (\%) & Mapped domain name & Good & Meaningful\\
    \hline
    0 & dropbox.com & Yes & No \\
    10 & dropbox.com & Yes & No \\
    20 & dropbox.com & Yes & No \\
    25 & apple.com & Yes&Yes \\
    30 & apple.com & Yes&Yes \\
    40 & apple.com & Yes& Yes \\
    50 & apple.com & Yes & Yes\\
    \hline
    \end{tabular}
    \caption{Clustering result for \texttt{Client C} with varying percentages of indicative connections among 300 connections}
    \label{tab:numindict}
\end{table*}


\subsection{Test on 10k TLS Connections}
We ran \name on 10,000 TLS connections from our data set to test the tool's robustness. This evaluation was done on a c3-highmem-8 machine with 8 vCPUs and 32GB of memory,  on Intel Sapphire Rapids. For this experiment, both \texttt{Z} and \texttt{H} values were set to 1, as this is the most basic setting for \name. We observed that \name maps more than 90\% of the client clusters to exactly one domain cluster whose weight has z-score larger than 1. 


\subsection{Computational Cost}
\name does not scale efficiently for large numbers of TLS connections. As shown in figure 5 its memory footprint increases faster than linearly as the number of TLS connections increases. All tests were done using a c3-standard-88 machine with 88 vCPU and 352GB of memory. It is projected to take at least 350GB of memory to process 50,000 TLS connections. Therefore, if a user wishes to process a large set of TLS connections, we recommend dividing them into batches of approximate size and running \name on them separately. Connections made by the same TCP fingerprints have to be included in the same batch to ensure a proper mapping.

\begin{figure}[h] \label{memory}
\begin{center}
\fbox{\includegraphics[width=0.4\paperwidth]{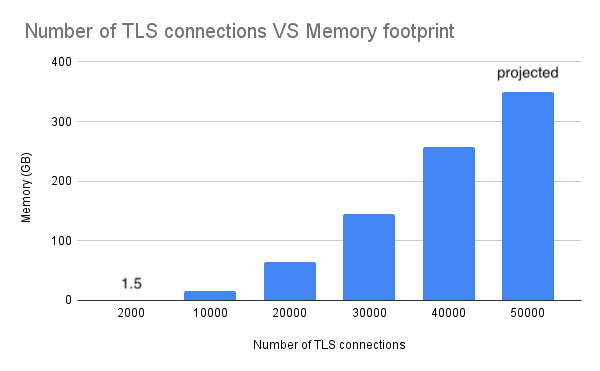}}
\caption{Memory footprint by number of TLS connections}
\end{center}
\end{figure}
\section{Related Work}

There are multiple client identification tools that build on previously established databases. P40f \cite{princeton} is a tool that uses p0f to fingerprints the OS of hosts running in their network and additionally react to it (e.g drop host). However, p0f is able to classify only 23\% of their collected traffic.

Using SNI as a window into network clients' identity is not new \cite{sni, sni1}. Some solutions have created a dictionary of chosen domain names that they have tested to be associated with certain OSes \cite{Lastovicka}. However, a known SNI-OS database can classify only 5.16\% of all unique TCP fingerprints in our TLS database, which consists of real traffic observed in the Stanford University campus network. While we share the insight of such database-based approaches that the frequency, type, or hosts of particular SNIs can be indicative of clients' identities, \name differs from them in that it can generate new matchings between clients and SNI unique to each input set of TLS connections. Given that no client identification databases is updated automatically, we believe that \name is able to process clients on the most up-to-date information. 

Furthermore, there are some client identification tools that directly infer the OS from user-agent available in HTTP headers \cite{ua2, ua3}. However, because more than 95\% of all network over Google as of May 2023 \cite{google} use some network encryption like HTTPS, payload-based identification like using user-agent is mostly unavailable. There also exist approaches that use user-agent as ground truth \cite{ua1} to verify models that use other features, but it has been known that user-agent can be easily fabricated. Unlike these approaches, the SNI included in the client hello of TLS connections are difficult to be tempered with and therefore provide more reliable information about the clients.


\section{Discussion} \label{sec:discussion}




A limitation of \name is that it does not entirely automate the mentioned strawman client identification method, in which a human inspects all domain names accessed by a client and tries to infer its identity from them. After \name outputs a domain cluster, a person still needs to manually go through the domain names, which sometimes requires connecting to the sites, and understand what they may indicate about the client. While \name reduces the number of domain names that a person would need to inspect this way by selecting the most 'useful' domain names for them, it does not completely replace the manual effort. Therefore, two desirable improvements from \name would be (1) establishing a more reliable way to know what a domain names represent and (2) automating the domain name lookup process. We envision there to be modern solutions to achieve these goals, such as feeding the domain names into Large Language Models like ChatGPT and retrieving a summary of what the sites most likely indicate about a frequent client.
\section{Conclusion}
In this paper, we examined the status quo of passive client identification by testing the performance of some of the most well-known client identification databases in a real-world network. We showed that the current rule-based approaches fail to identify many TLS clients and proposed \name, which uses unsupervised learning on the SNI of TLS connections to match clusters of clients to clusters of domain names. Using a weight formula based on the frequency and exclusivity of connections, \name is able to identify the most associated domain names for at least 60\% of client clusters in any given set of TLS connections. While there are still unresolved challenges arising from the lack of ground truth in clients' identities in real-world networks, \name offers a promising alternative to current rule-based fingerprinting tools.

However, more work is required to verify the usefulness of \name's identification results. Doing so is non-trivial because \name is a general purpose identification tool that attempts to pinpoint domain names that will \textit{likely} tell us something about the clients. While we believe that SNI is a strong indicator of certain aspects of clients, \name is not guaranteed to identify definitive features, like the operating system, manufacturer, or the type of device of clients. Therefore, the usefulness of the extracted information depends on how telling the domain names that the client accessed are and what a human user of \name get out of them. Interpreting the domain names selected by \name could be made easier by using large language models like ChatGPT to automatically connect to the domain and tell the user what the site suggests about, for example, the operatin system about the client device.

\section*{Acknowledgement}
This work was supported in part by the National Science Foundation under awards CNS-2124424 and IMR-2319080, and through gifts from Google, Inc., Cisco Systems, Inc., and Comcast Corporation. We also thank the members of Stanford's Empirical Security Research Group for their valuable feedback on this work.
{\footnotesize \bibliographystyle{acm}
\bibliography{ref}}

\end{document}